\documentclass[twocolumn,showpacs,prl]{revtex4}

\usepackage{amsmath,amssymb}
\usepackage{graphicx}
\usepackage{verbatim}

\draft
\begin{document}

\title{Quantum Spin Current Induced Through Optical Fields}

\author{Xiong-Jun Liu$^{a}$\footnote{Electronic address:
phylx@nus.edu.sg}, Xin Liu$^{b,c}$, L. C. Kwek$^{a,d}$, C. H.
Oh$^{a}$\footnote{Electronic address: phyohch@nus.edu.sg} and
Mo-Lin Ge$^{b}$} \affiliation{a. Department of Physics, National
University of Singapore,
2 Science Drive 3, Singapore 117542 \\
b. Theoretical Physics Division, Nankai Institute of
Mathematics,Nankai University, Tianjin 300071, P.R.China\\
c. Department of Physics, Texas A\&M University, College Station,
Texas 77843-4242, USA\\
d. National Institute of Education, Nanyang Technological
University, 1 Nanyang Walk, Singapore 639798}

\begin{abstract}
We propose a scheme to generate quantum spin current via optical
dipole transition process. By coupling a three-level system based
on the spin states of charged particles (electrons or holes in
semiconductor) to the angular momentum states of the radiation, we
show that a pure quantum spin current can be generated. No
spin-orbit interaction is needed in this scheme. We also calculate
the effect of nonmagnetic impurities on the created spin currents
and show that the vertex correction of the spin hall conductivity
in the ladder approximation is exactly zero.
\end{abstract}
\date{\today }
\maketitle

\indent Spintronics \cite{spintronics1,spintronics2}, the science
and technology of manipulating the spin of the electron for
building integrated information processing and storage devices,
showed great promise and developed rapidly in recent years. In
practical application, one of the most important goals is to
create spin currents \cite{zhang,niu}. For this many interesting
and basic phenomena, e.g. the spin hall effect
\cite{niu,hall1,hall2} in the system with spin-orbit coupling,
have been discovered and further studied. On the other hand, spin
currents can also be generated with the interference of two
optical fields \cite{optical1,optical2,optical3}, or with optical
Raman scattering effects \cite{sipe}. Most of these methods in the
generating quantum spin currents rely on the spin-orbit coupling
such as Rashba coupling (caused by asymmetry of quantum well) and
Dresselhaus term (induced by asymmetry of bulk crystal), etc.
Moreover, coherent manipulation of spin states in semiconductor
via optical dipole transitions has theoretically and
experimentally been studied by many authors
\cite{coherence1,coherence2}. With these ideas in mind, we propose
here a new scheme to generate spin currents.

In this letter, we consider a realization of spin currents via
optical dipole transition process in three-level system based on
the spin states of the electrons (n-doped) or holes (p-doped) in
semiconductor. The short-range scattering by the nonmagnetic
impurities is discussed and the vertex correction of the spin hall
conductivity is shown to be zero in the present optical model.

\begin{figure}[ht]
\includegraphics[width=0.7\columnwidth]{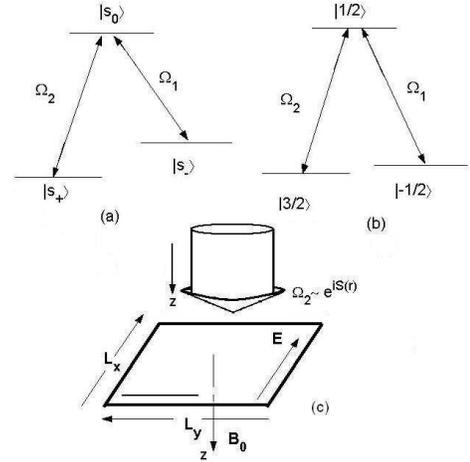}
\caption{(a)$\Lambda$-type system based on spin states. (b) An
example for dipole transitions in the $\Lambda$-type system
provided by semiconductor GaAs quantum well, where the excited
state $|1/2\rangle$ is a conduction band state, and the two ground
states correspond to light-hole state (for $|-1/2\rangle$) and
heavy-hole state (for $|3/2\rangle$), see ref.\cite{nonlinear} and
references therein. (c) Illustration of the two-dimensional system
for present model.} \label{fig1}\end{figure}

The three-level $\Lambda$-type system that we envisage for the
scheme is given in Fig. \ref{fig1}. The system is confined to the
two-dimensional $x-y$ plane with area $L_x\times L_y$ provided by
the semiconductor quantum well. The particles in the system are
subjected to an uniform electric field $\bold E$ in the $+x$
direction and a $+z$-directional uniform magnetic field ($\bold
B_0=B_0\hat{\bold e}_z$) which can cause energy splitting and spin
polarization in the $z$ direction ($S_z=s_0,s_+,s_-$, see
Fig.1(a)). The transition $|s_-\rangle\rightarrow|s_0\rangle$ is
coupled by a $\sigma_+$ light with the Rabi-frequency
$\Omega_1=\Omega_1^{(0)}\exp(i\bold k_1\cdot\bold r)$, where
$\Omega^{(0)}_1(r)$ is the slowly spatially varying amplitude and
$\bold k_1$ is the wave-vector. Another $\sigma_-$ light,
characterized by the Rabi frequency
$\Omega_2=\Omega^{(0)}_2(r)e^{i(l\phi+\bold k_2\cdot\bold r)}$
with $\Omega^{(0)}_2(r)$ the slowly spatially varying amplitude,
couples the transition $|s_+\rangle\rightarrow|s_0\rangle$, where
$\phi=\tan^{-1}({y}/{x})$ and $\bold k_2$ is the wave-vector. $l$
indicates that the $\sigma_-$ photons are assumed to have the
orbital angular momentum $\hbar l$ along the $+z$ direction.
 As in previous works
\cite{zhang,niu,hall2,nonlinear}, we focus on the generation of
spin current without particle-particle interaction. Defining the
flip operators as $\hat\sigma_{\mu\nu}=|\mu\rangle\langle\nu|$
with energy levels $\mu,\nu=s_0,s_+,s_-$, the Hamiltonian in
present system reads
\begin{eqnarray}\label{eqn:H1}
H&=&H_0+H_I\\
H_0&=&\frac{1}{2m_e}(i\hbar\nabla_{xy}+e{\bold A}_0)^2+V(x)-g_s\mu_BB_0S_z\nonumber\\
H_I&=&-(\hbar\Omega_2\hat\sigma_{s_0s_+}+\hbar\Omega_1\hat\sigma_{s_0s_-})+h.c.\nonumber,
\end{eqnarray}
where $\nabla_{xy}=\hat{\bold e}_x\partial_x+\hat{\bold
e}_y\partial_y$, $V(x)=-eEx$ is the electric potential, parameters
$e$, $m_e$, $\mu_B$ and $g_s$ represent the charge, the effective
mass of a particle, Bohr magneton and Lande factor, respectively.
${\bold A}_0$ is the vector potential of the applied magnetic
field, i.e. $B_0\hat{\bold e}_z=\nabla\times{\bold A}_0$. Note
that spin-orbit interaction is not required in our model (we
consider a very small or zero Rashba term in present case, see
e.g. \cite{Rashba}). To facilitate the discussion, $H_0$ is
written in $\bold r$-representation in eq (\ref{eqn:H1}). The
interaction Hamiltonian $H_I$ can be diagonalized with the local
unitary transformation: $\tilde{H}_I=U(\bold r)H_IU^{\dag}(\bold
r)$ with
\begin{eqnarray}\label{eqn:unit1}
U(\bold r)={\left[ \begin{matrix} \frac{1}{\sqrt{2}}&
\frac{1}{\sqrt{2}}
\sin\theta e^{-iS(\bold r)} & \frac{1}{\sqrt{2}}\cos\theta\\
0& \cos\theta& -\sin\theta e^{iS(\bold r)}\\
 \frac{1}{\sqrt{2}}& -\frac{1}{\sqrt{2}}\sin\theta e^{-iS(\bold r)} &
 -\frac{1}{\sqrt{2}}\cos\theta& \end{matrix} \right]},
\end{eqnarray}
where $S=(\bold k_{2}-\bold k_{1})\cdot \bold r+l\phi$ and the
mixing angle $\tan\theta(r)=|\Omega_2/\Omega_1|$. Under this
unitary transformation, the three eigenstates of $H_I$ are easily
obtained as $
|\Psi_{\alpha}\rangle=\sum_{j=0,\pm}w^{(i)}_j|s_j\rangle$, where
$\alpha=0,\pm$). The coefficients of these states are
$w^{(0)}_0=0$, $w^{(0)}_+=\cos\theta$, $w^{(0)}_-=-\sin\theta$,
$w^{(\pm)}_{0}=\frac{1}{\sqrt{2}},
w^{(\pm)}_{+}=\pm\frac{1}{\sqrt{2}}\sin\theta$ and
$w^{(\pm)}_{-}=\pm\frac{1}{\sqrt{2}}\cos\theta$, and the
corresponding eigenvalues read $E^{(0)}=0,
E^{(\pm)}=\pm\sqrt{|\Omega_2|^2+|\Omega_1|^2}$. Eigenstate
$|\Psi_0\rangle$ is typically known as a dark state for a
three-level $\Lambda$ system \cite{dark-state}.

The Hamiltonian (\ref{eqn:H1}) can be rewritten in a covariant
form with the definition of covariant derivative operator
$\hat{\bold D}=-i\nabla-\bold{\tilde{A}}$, where
$\bold{\tilde{A}}={\bold A}_0+{\bold A}_e$ is the sum of Abelian
gauge potential ${\bold A}$ and the non-Abelian one ${\bold
A}_e=-i U(\bold r)\nabla U^{\dag}(\bold r)$, and the latter
definition does not lead to the curvature term since it is a pure
gauge. By a straightforward calculation, we find the $3\times3$
matrix $\bold{\tilde{A}}$ has the following form:
\begin{widetext}
\begin{eqnarray}\label{eqn:gauge1}
{\left[ \begin{matrix} \frac{1}{\sqrt{2}}\sin^2\theta\nabla
S+{\bold A}_0&
(i\frac{1}{\sqrt{2}}\nabla\theta+\frac{1}{2\sqrt{2}}
\sin2\theta\nabla S)e^{-iS} & -\frac{1}{\sqrt{2}}\sin^2\theta\nabla S\\
(-i\frac{1}{\sqrt{2}}\nabla\theta+\frac{1}{2\sqrt{2}} \sin2\theta
\nabla S)e^{iS}& -\sin^2\theta\nabla S+{\bold A}_0&
(i\frac{1}{\sqrt{2}}\nabla\theta-\frac{1}{2\sqrt{2}}
\sin2\theta\nabla S)e^{iS}\\
-\frac{1}{\sqrt{2}}\sin^2\theta\nabla S&
(-i\frac{1}{\sqrt{2}}\nabla\theta-\frac{1}{2\sqrt{2}} \sin2\theta
\nabla S)e^{-iS} & \frac{1}{\sqrt{2}}\sin^2\theta\nabla S+{\bold
A}_0\end{matrix} \right]}
\end{eqnarray}
\end{widetext}
which generally is a $SU(3)$ non-Abelian gauge potential. The
off-diagonal matrix elements of it represent the transitions
between each two of $|\Psi_0\rangle$ and $|\Psi_{\pm}\rangle$.
However, considering the present system is non-degenerate, we can
introduce the adiabatic approximation when both light fields vary
sufficiently slowly in space. For a numerical estimate, we
consider the case
$|\frac{\Omega_2}{\Omega_1}|=\lambda\rho\propto\rho$ where
$\lambda$ is constant and $\rho=\sqrt{x^2+y^2}$ the distance from
$z$ axis \cite{ohberg}. The typical values can be taken as
\cite{coherence1}:
$\Omega_0^2=(\Omega_1^2+\Omega_2^2)^{1/2}\sim10ps^{-1}$,
$l\leq10^4$, $\lambda\sim10^4m^{-1}$ and the radius of the
interaction region in the $x-y$ plane is $R\sim1.0cm$, the
transition rate can then be evaluated as $\Gamma_{ab}<10^{-3}\ll1$
for any two different states ($a\neq b=\pm,0$). Thus we neglect
the off-diagonal matrix elements in eq.(\ref{eqn:gauge1}) and
yield a nontrivial adiabatic gauge connection
$\bold{\tilde{A}}\rightarrow\bold{\tilde{A}^{diag}}$, which is
associated with a $3\times3$ diagonal matrix for the magnetic
field with each component
($B_k=\frac{1}{2}\epsilon_{kmn}F_{mn}=\frac{\hbar}{2}\epsilon_{kmn}[\hat
D_m, \hat D_n]$)
\begin{eqnarray}\label{eqn:magnetic}
\bold B_{11}=\bold B_{33}=(B_0+B_e)\hat{\bold e}_z, \ \ \ \bold
B_{22}=(B_0-2B_e)\hat{\bold e}_z
\end{eqnarray}
where $B_e=\hbar l\lambda^2/e$ indicates the strength of
additional effective magnetic fields induced by the optical fields
depends on the angular momentum $\hbar l$ that can have a large
value from a vortex optical beam \cite{angular}. For our present
purpose we choose $B_0=B_e/2$ so that
$B_{11}^z=B_{33}^z=-B_{22}^z=3\hbar l\lambda^2/2e$. In this case,
we obtain an interesting result that the $z$ component of $\bold
B_{22}$ is opposite to that of $\bold B_{11}$ (and of $\bold
B_{33}$) (see fig.2(a) and (b)), a key point for realizing quantum
spin current.
\begin{figure}[ht]
\includegraphics[width=0.8\columnwidth]{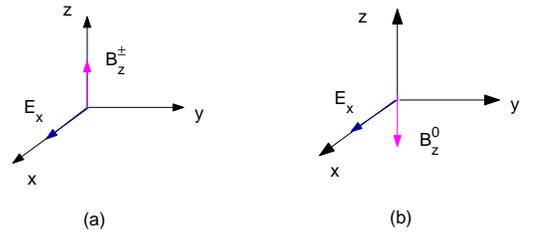}
\caption{(color online) (a)(b) Schematic illustration of the
electric field and the effective magnetic field, where $\bold
B^0=\bold B_{22}, \bold B^+=\bold B^{11}$ and $\bold B^-=\bold
B^{33}$.} \label{}
\end{figure}

Up to this point, the interaction Hamiltonian is diagonal and we
can rewrite $H$ in eq.(\ref{eqn:H1}) in the following effective
form:
\begin{eqnarray}\label{eqn:H2}
H_{eff}=\frac{\hbar^2}{2m_e}\hat D^2+\tilde{V}(\bold r),
\end{eqnarray}
where $\tilde{V}(\bold r)=\bigr(U(\bold r)VU(\bold
r)\bigr)^{diag}$ is a $3\times3$ diagonal matrix with its elements
$\tilde{V}_{11}=\tilde{V}_{22}+\Omega_0,
\tilde{V}_{22}=V(x)-g_s\mu_BB_0S_z$ and
$\tilde{V}_{11}=\tilde{V}_{22}-\Omega_0$. The momentum operator is
defined by $\hat{\bold p}=-i\hbar\hat{\bold D}$ and it has the
nontrivial communication relations: $[r_i, \hat
p_j]=i\hbar\delta_{ij}, [r_i, r_j]=0, [\hat p_i, \hat p_j]=-i\hbar
F_{ij}$. Generally the dynamics of this situation is determined by
the $SU(3)$ gauge symmetry. By considering adiabatic condition,
the transitions between any two of the states
$|\Psi_{\alpha}\rangle (\alpha=0,\pm)$ can be neglected and the
symmetry $SU(3)$ readily reduces to an Abelian one:
$SU(3)\rightarrow U(1)\otimes U(1)\otimes U(1)$. The quantum state
of each particle can be expanded by the complete eigenbasis of the
interaction Hamiltonian
$|\bold\Psi\rangle=\sum_{\alpha=0,\pm}\Phi_{\alpha}|\Psi_{\alpha}\rangle$,
where the coefficients $\Phi_{\alpha}$ can be determined by the
initial conditions. Note that each eigenstate
$|\Psi_{\alpha}\rangle$ consists of different spin states
$|s_0\rangle$ and $|s_{\pm}\rangle$. However, since
$|\Psi_{\alpha}\rangle$ are eigenstates of the interaction
Hamiltonian $H_I$, and $H_0$ does not lead to the transition
between $|s_0\rangle$ and $|s_{\pm}\rangle$, the expectation value
of the $z$-axis spin polarization of any state
$|\Psi_{\alpha}\rangle$ is time-independent. Therefore, for
convenience, we can treat the states $|\Psi_{\alpha}\rangle$
approximately as new ``effective spin states" with their $z$-axis
spin polarization calculated by
$S_z^{\alpha}=\langle\Psi_{\alpha}|S_z|\Psi_{\alpha}\rangle$. It
is easy to see that $S_z^{(0)}=\cos^2\theta s_++\sin^2\theta s_-$,
$S_z^{(+)}=S_z^{(-)}=\frac{1}{2}s_0+\frac{1}{2}(\sin^2\theta
s_++\cos^2\theta s_-)$. Since the particles in different effective
spin states $|\Psi_{\alpha}\rangle$ experience different effective
magnetic fields, they may move in opposite directions depending on
the index $\alpha$, generating a spin current.

To facilitate the subsequent calculations, we note that the
particles with effective spin state $|\Psi_{\alpha}\rangle
(\alpha=0,\pm)$ interact with the magnetic field $\bold
B^{\alpha}$ where $\bold B^0=\bold B_{22}$ and $\bold
B^{\pm}=\bold B_{11}=\bold B_{33}$. It is convenient to choose the
diagonal-elements of the gauge connection $\bold{\tilde{A}}(\bold
r)$ as
$\bold{\tilde{A}}_{\alpha}=\frac{e}{c}B^{\alpha}_zx\hat{e}_y$, so
that $p^{\alpha}_y=k$ is a good quantum number. With the
definition $\hat a_{\alpha
k}=\frac{1}{\sqrt{2}l_B}\bigr[(x-\frac{Ec}{B\omega})
+(-\frac{k}{B^{\alpha}}+i\frac{p_x}{B})\frac{c}{e}\bigr]$ where
$B=|B^{\alpha}|=3B_e/2$ and $l_B=\sqrt{\hbar c/eB}$, the
Hamiltonian for a given $k$ can further be rewritten as
\begin{eqnarray}\label{eqn:H3}
H_{eff}=\hbar\omega\bigr(\hat a^{\dag}_{\alpha k}\hat a_{\alpha
k}+\frac{1}{2}\bigr)-gB_0S_z^{\alpha}+H_c,
\end{eqnarray}
where $H_c=-\frac{E_x^2}{2B^2}mc^2-\frac{E_x}{B^{\alpha}_z}kc$,
$g=g_s\mu_B$ and $\omega=\frac{eB}{m_ec}$. Operators $\hat
a_{\alpha k}$ satisfy the communication relation $[\hat a_{\alpha
k},a^{\dag}_{\beta k'}]=\delta_{\alpha\beta}\delta_{kk'}$. The
eigenstate of above Hamiltonian can be written as
$|n,k,\alpha\rangle$ with its eigenvalue
$\epsilon_{n,k,\alpha}=(n+1/2)\hbar\omega-gB_0S_z^{\alpha}+H_c$.
The analytical results allow us to calculate the charge and spin
current. The spin current operator for a single particle is
defined by $j^{s_z}_s=\frac{\hbar}{2}(S_zv_y+v_yS_z)$, where
$v_y=[y,H]/i\hbar=p_y+x\omega$ is the velocity operator in the $y$
direction \cite{hall2} and the corresponding charge current
operator reads $j_c=ev_y$.

With the above definitions, for a $N_e$-particle system the
average current density can be calculated by
\begin{eqnarray}\label{eqn:current2}
J^{s_z}_{c,y}=\frac{1}{L_x}\sum_{n,k,\alpha}(j^{s_z}_{c,y})_{nk}f(\epsilon_{n,k,\alpha}).
\end{eqnarray}
where $(j^{s_z}_{c,y})_{nk}=\sum_{\alpha}\langle
n,k,\alpha|j^{s_z}_{c,y}|n,k,\alpha\rangle$ is the current carried
by one particle in the state $|n,k\rangle$.
$f(\epsilon_{n,k,\alpha})$ is the Fermi distribution function, and
$N_e=\sum_{n,k,\alpha}f(\epsilon_{n,k,\alpha})$. When the
coefficient $\Phi_{\alpha}$ of the initial state takes on the
simple value $|\Phi_0|^2=1/2$, $|\Phi_{+}|^2=\cos^2\gamma/2$ and
$|\Phi_{-}|^2=\sin^2\gamma/2$, a pure spin current is obtained by
\begin{eqnarray}\label{eqn:spincurrent2}
(j^{s_z}_y)_{nk}=\frac{eE_x}{6\hbar
l\lambda^2}\bigr(\cos2\theta(s_+-s_-)-s_0\bigr),\ \ j_c=0.
\end{eqnarray}
The spin current in above equation is dependent on the space
position $(x,y)$. This is because the effective spin polarization
of the state $|\Psi_{\alpha}\rangle$ is dependent on space (see
Fig. 3). Together with the Eq. (\ref{eqn:current2}) we further
calculate the average spin current \begin{figure}[ht]
\includegraphics[width=1.0\columnwidth]{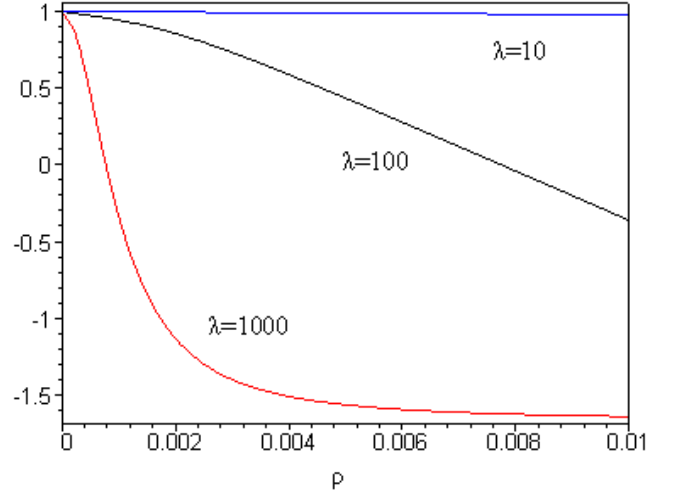}
\caption{(color online) Space spreading of generated spin current
versus the unit $\frac{eE_x}{9\hbar l\lambda^2}$, where
$\lambda=10 (1/m)$ for the blue line, $\lambda=100 (1/m)$ for the
black line and $\lambda=1000 (1/m)$ for the red line.
$\rho=\sqrt{x^2+y^2}$ versus the unit $m$.} \label{}\end{figure}
\begin{eqnarray}\label{eqn:average1}
J^{s_z}_y=\frac{eE_xn_e}{6\hbar
l\lambda^2}\bigl(4s_{\pm}\frac{\tan^{-1}\bigr(\lambda
L_x/2\sqrt{1+\lambda^2y^2}\bigr)}{\lambda
L_x\sqrt{1+\lambda^2y^2}}-s_{\pm}-s_0\bigr),
\end{eqnarray}
where $s_{\pm}=s_+-s_-$, and $n_e$ is the filling of charged
particles (unit $1/m^2$). It is interesting that generally the
average spin current is still dependent on the position $y$. For
practical application, here we consider the case that the $\lambda
L_x\gg1$, i.e. the $\sigma_-$ field is much stronger than
$\sigma_+$ one. Then from the Eq. (\ref{eqn:average1}) we obtain
the persistent spin current $ J^{s_z}_y=-\frac{eE_xn_e}{6\hbar
l\lambda^2}(s_{\pm}+s_0)$. To provide some numerical evaluations,
we set \cite{hall2,coherence1} $n_e=1.0\times10^{16}/m^2,
l\lambda^2=3.0\times10^{13}/m^2$ ($\lambda L_x\gg1$) and the spin
polarization $s_+=3\hbar/2, s_-=-\hbar/2, s_0=\hbar/2$ (fig.1(b)).
For these values, we obtain the spin hall conductivity
$\sigma_{yx}^{s}=J^{s_z}_y/E_x\approx1.38\times10^{2}$e. The above
result can be understood as follow: from the formula of $H_c$ one
can find the particles in the effective spin state
$|\Psi_{\alpha}\rangle$ have a group velocity $\bold
v^{\alpha}=-E_x/B^{\alpha}_z\hat{e}_y$ in the $y$ direction. As in
present case $B^+=B^-=-B^0$, the group velocities $v^+=v^-=-v^0$.
Therefore, a pure dissipationless spin current can be generated
when the particles have the same probability in state
$|\Psi_0\rangle$ with the sum of $|\Psi_+\rangle$ and
$|\Psi_-\rangle$. It should be pointed out that, for GaAs quantum
well, this numerical evaluation should be revised by a constant
factor, because another independent three-level system
($s'_+=\hbar/2, s'_-=-3\hbar/2, s'_0=-\hbar/2$, noted by
$\Lambda'$ system) can also interact with the optical fields
besides the one shown in fig.1(b) \cite{nonlinear1}. If we neglect
the detunings of the optical transitions and according to
$s'_{\pm}+s'_0=3\hbar/2<s_{\pm}+s_0$, the $\Lambda'$ system will
slightly decrease the present generated spin currents.

By now all the discussions are based on a clean semiconductor
system. Noting that the vertex correction may remarkably cancel
the spin hall currents in Rashba system when the s-wave impurity
scattering is considered \cite{vertex1,vertex2}, it is much
suggestive to discuss such effect in the present model. For this
we consider the following short-range nonmagnetic impurity
Hamiltonian:
\begin{eqnarray}\label{eqn:impurity1}
H_{imp}=\sum_{j}u\delta(\bold r-\bold R_j).
\end{eqnarray}
The free retarded Green's function $G_R^{(0)}(\epsilon,
k)=[\epsilon-H_{eff}]^{-1}$ is diagonal in the effective spin
space:
\begin{eqnarray}\label{eqn:green1}
G_{R;n,\alpha}^{(0)}(\epsilon,
k)=\frac{1}{\epsilon-\epsilon_{n,k,\alpha}+i\eta}\delta_{nn'}\delta_{\alpha\alpha'}.
\end{eqnarray}
In the Born approximation, the self-energy related to free green
function in our system can be calculated by
$\Sigma^R(\epsilon)=n_{imp}u^2\int
\mathcal{N}_e(\epsilon_{n,k,\alpha})G_{R;n,\alpha}^{(0)}(\epsilon,
k)d\epsilon_{n,k,\alpha}$, where $n_{imp}$ is the impurity density
and $\mathcal{N}_e(\epsilon_{n,k,\alpha})$ the electron state
density. It can be verified that $\Sigma^R(\epsilon)=-i\pi
n_{imp}u^2\mathcal{N}_e(\epsilon)$, thus the total Green's
function reads $G_{R;n,\alpha}(\epsilon,
k)=\frac{1}{\epsilon-\epsilon_{n,k,\alpha}+i/2\tau}\delta_{nn'}\delta_{\alpha\alpha'}$.
Here $\tau=2\pi n_{imp}u^2\mathcal{N}_e$ represents the mean
scattering free time. With these results we calculate the vertex
correction of charge current $j_{\alpha,x}^{vertex}$ in the ladder
approximation as
\begin{widetext}
\begin{eqnarray}\label{eqn:green2}
\sum_{n,\alpha}\frac{1}{2\pi\tau
\mathcal{N}_e}\int\frac{dk}{2\pi}G_{A;n,\alpha}(\epsilon,
k)j_{\alpha,x}G_{R;n,\alpha}(\epsilon,
k)&=&\sum_{n,\alpha}\frac{e}{2\pi m\tau
\mathcal{N}_e}\biggr(\int\frac{dk}{2\pi}G_{A;n,\alpha}(\epsilon,
k)\sqrt{\frac{eB\hbar}{2c}}\bigr(\sqrt{n+1}\delta_{n,n'-1}+\nonumber\\
&&+\sqrt{n}\delta_{n,n'+1}\bigr)G_{R;n,\alpha}(\epsilon,
k)\biggr),
\end{eqnarray}
\end{widetext}
where the advanced Green's function $G_{A;n,\alpha}(\epsilon,
k)=\{G_{R;n,\alpha}(\epsilon, k)\}^*$. Notice that the Green's
function in the present system is diagonal in both Landau level
and effective spin space: $G^{A,R}_{n,\alpha}(\epsilon,
k)\propto\delta_{nn'}\delta_{\alpha\alpha'}$, whereas integral
function in the right-hand side of above equation is also
proportional to $\delta_{n,n'+1}$ or $\delta_{n,n'-1}$, the vertex
correction of charge current in ladder approximation is exactly
zero. When we turn to the ladder diagrams of the spin Hall
conductivity $\sigma_{yx}^s$, we have to take into account the
vertex correction in the charge current only \cite{vertex1}.
Therefore, because of this vanishing vertex correction, the spin
current obtained in our optical method reproduces its value
obtained from a clean system without impurities from the outset.

Finally, we present an experimental realization of the initial
condition required for the result in eq. (\ref{eqn:current2}).
Firstly, we can trap the system in dark state using two light
fields with equal Rabi-frequencies (i.e. $|\Omega_1|=|\Omega_2|$)
and switch both lights off simultaneously, hence the state
$|\Psi(t_0)\rangle=(|s_+\rangle-|s_-\rangle)/\sqrt{2}$. In this
step no angular momentum is needed for the $\sigma_-$ light field.
Secondly, we only turn the $\sigma_+$ light on again. Meanwhile
the population in the state $|s_+\rangle$ keeps $1/2$ unchanged
while the state $|s_-\rangle$ is coupled to the state
$|s_0\rangle$ by the $\sigma_+$ light and the state reads
$|\Psi(t_1)\rangle=(\sin|\Omega_1|(t_1-t_0)|s_0\rangle+
\cos|\Omega_1|(t_1-t_0)|s_-\rangle+|s_+\rangle/)\sqrt{2}$.
Thirdly, we adiabatically turn on the $\sigma_-$ light field with
$z$-directional angular momentum and arrive at the adiabatic
evolution of the state
$|\Psi(t)\rangle=\frac{1}{\sqrt{2}}\bigr(\frac{1}{\sqrt{2}}e^{-i\gamma
t}|\Psi_+\rangle -\frac{1}{\sqrt{2}}e^{i\gamma
t}|\Psi_-\rangle+|\Psi_0\rangle\bigr)$ where
$\gamma=\int_{t_0}^{t}\sqrt{|\Omega_1|^2+|\Omega_2|^2}dt$.
Obviously, this is the state needed for the initial condition.

In practice, the spreading of light fields have a boundary in the
$x$-$y$ plane which may lead to modification in the Landau energy
levels. However, based on the previous results \cite{ohberg}, this
boundary effect can be safely neglected when
$l\lambda^2L_xL_y\gg1$, which is achieved using a light beam with
a large angular momentum. On the other hand, we should emphasize
that the particle-particle interaction may lead to a
renormalization of the Rabi-frequencies $\Omega_{1,2}$ of the
transitions between states $|s_0\rangle$ and $|s_{\pm}\rangle$
coupled by the light fields \cite{nonlinear1,nonlinear2}. However
for simplicity, as in previous works
\cite{zhang,niu,hall2,nonlinear}, we do not need to consider such
effects in the present model.  All these interesting aspects will
be further discussed in future publications. In conclusion, we
have proposed and demonstrated a means of generating quantum spin
current optical dipole transition process. The short-range
scattering by the nonmagnetic impurities are discussed and the
vertex correction of the spin hall conductivity is shown to be
zero in the present optical model.

\bigskip

We thank Prof Mansoor B. A. Jalil, S. -Q. Gong and Dr S. G. Tan
for valuable discussions. This work is supported by NUS academic
research Grant No. WBS: R-144-000-172-101, and by NSF of China
under grants No.10275036.


\noindent\\

\end{document}